\documentclass[12pt]{article}
\usepackage{amssymb,amsmath,epsfig}

\begin{document}
\title{\bf Reconstructing QCD ghost $f(R,T)$ models}

\author{M. Zubair$^1$ \thanks{mzubairkk@gmail.com; drmzubair@ciitlahore.edu.pk}
and G. Abbas$^2$ \thanks{ghulamabbas@ciitsahiwal.edu.pk}\\\\
$^1$ Department of Mathematics, COMSATS\\
Institute of Information Technology, Lahore-54000, Pakistan.\\
$^2$ Department of Mathematics, COMSATS\\
Institute of Information Technology, Sahiwal, Pakistan.}

\date{}

\maketitle

\begin{abstract}
We reconstruct $f(R,T)$ theory (where $R$ is the scalar curvature and $T$ is
the trace of energy-momentum tensor) in the framework of QCD ghost dark
energy models. In this study, we concentrate on particular models of $f(R,T)$
gravity which permits the standard continuity equation in this theory. It is
found that reconstructed function can represent phantom and quintessence
regimes of the universe in the background of flat FRW universe. In addition,
we explore the stability of ghost $f(R,T)$ models.
\end{abstract}

\section{Introduction}

Contemporary observational results form Supernovae type Ia (SNeIa)
(Perlmutter et al. 1999) revealed the expanding behavior of the
universe. This fact has further been affirmed by the observations of
anisotropies in cosmic microwave background (CMB) (Spergel et al.
2004)), large scale structure (Hawkins, E. et al. 2003), baryon
acoustic oscillations (Eisentein, D.J. et al. 2005) and weak lensing
(Jain and Taylor 2003). The most promising feature of the universe
is the dominance of exotic energy component with large negative
pressure, known as \emph{dark energy} (DE). A number of alternative
models have been proposed in the framework of general relativity
(GR) to explain the role of DE in the present cosmic acceleration
(Bamba et al. 2012).

Recently, a new dynamical DE model is proposed on the basis of Veneziano
ghost choromodynamics (QCD) called as ghost DE (GDE) (Urban and Zhitnitsky
2009; Ohta 2011). The existence of Veneziano ghost is necessary for the
resolution of $U(1)$ problem in low energy effective theory (Witten 1979;
Veneziano 1979). Although in the usual Minkowski spacetime QFT, Veneziano
ghost is unphysical and makes no contribution to the vacuum energy denisty
but it exhibits significant physical effects in dynamical spacetime. In a
curved spacetime this ghost provides the vacuum energy density proportional
to $\Lambda^3_{QCD}H$ (Zhitnitsky 2010), where $H$ is the Hubble parameter
and $\Lambda_{QCD}\sim100$Mev is the QCD mass scale. For $H\sim10^{-33}$eV,
$\Lambda^3_{QCD}H$ gives the numerical value in accordance with the observed
energy density of DE. Therefore, GDE helps to get rid of fine tuning and
coincidence problems (Urban and Zhitnitsky 2009; Ohta 2011). Cai et al.
(2011) fitted this model and developed the constraints on model parameters
using the recent observational data SNeIa, CMB, BAO, BNN and the Hubble
parameter data.

GDE has attained significant attention and various aspects have been
discussed such as interacting GDE models (Sheykhi and Movahed 2012),
thermodynamics (Feng et al. 2012), correspondence with scalar field models
(Karami and Fahimi 2013), $f(R)$ gravity (Jawad 2014; Saaidi et al. 2012),
$f(\mathcal{T})$ gravity (Karami et al. 2013a) and Brans-Dicke theory
(Ebrahimi and Sheykhi 2011). In (Cai et al. 2012), authors presented a
generalized GDE of the form $\alpha{H}+\beta{H^2}$ and discussed its
dynamical evolution. The energy density of the QCD GDE can be related to the
radius of trapping horizon as (Garcia-Salcedo et al. 2013)
$\rho_{GDE}=\frac{\alpha(1-\epsilon)}{\tilde{r}_T}=\alpha(1-\epsilon)
\sqrt{H^2+\frac{\kappa}{a^2}}$, here
$\epsilon=\frac{\dot{\tilde{r}}_T}{2H\tilde{r}_T}$ introduces a new time
dependent component which is previously ignored. Garcia-Salcedo et al. (2013)
presented the phase space analysis for GDE and discussed some issues related
to the stability of this model. Jawad (2014) reconstructed the modified
Horava-Lifshitz $F(R)$ gravity for this QCD GDE model and discussed the
physical parameters.

The modification of Einstein-Hilbert action is another approach to
unravel the mysterious nature of DE and various candidates have been
proposed namely, $f(R)$ gravity (Capozziello and Faraoni 2011),
$f(\mathcal{T})$ (Ferraro and Fiorini 2007; Zubair 2015), where
$\mathcal{T}$ is the torsion scalar, Gauss–Bonnet gravity (De Felice
and Tsujikawa 2010), $f(R,T)$ gravity (Alvarenga et al. 2013a; Harko
et al. 2011; Sharif and Zubair 2012, Shabani and Farhoudi 2013;
Noureen and Zubair 2015, Noureen et al. 2015), where $T$ is the
trace of the energy-momentum tensor and
$f(R,T,R_{\mu\nu}T^{\mu\nu})$ gravity (Nojiri and Odintsov 2010;
2011a; Haghani et al. 2013; Sharif and Zubair 2013a, 2013b).
Cosmological reconstruction in modified theories is one of the
significant aspects in cosmology. The reconstruction schemes in
modified theories have been carried out under different scenarios
(Elizalde et al. 2004; Capozziello et al. 2006; Carloni et al. 2012)
to find out realistic cosmology which can explain the transition of
matter dominated epoch to DE phase. In (Nojiri and Odintsov 2006)
the general formulation of modified $f(R)$ gravity is presented
which can be reconstructed for FRW universe model. Several versions
of modified gravity are formulated compatible with solar system
tests which includes matter dominated phase, transition from
deceleration to acceleration, accelerating epoch and $\Lambda$CDM
cosmology consistent with recent observations. Nojiri and Odintsov
(2007, 2011b) discussed the reconstruction scheme in different
modified theories including scalar-tensor theory, $f(R)$, $f(G)$ and
string-inspired, scalar-Gauss-Bonnet gravity.

In this study, one interesting way is to consider the known cosmic evolution
and use the field equations to find particular form of Lagrangian that can
reproduce the given evolution background. The cosmological reconstruction has
been investigated in the framework of $f(R,T)$ gravity realizing the
$\Lambda$CDM, phantom, non-phantom eras and unification of matter dominated
and accelerated phases (Houndjo 2012; Sharif and Zubair 2013c, 2014a). In
this paper, we are interested to develop an equivalence between $f(R,T)$
gravity and GDE without utilizing any additional DE component. We reconstruct
the $f(R,T)$ model and discussed its evolution and stability. The paper is
arranged as follows. In next section we present the formulation of field
equations in $f(R,T)$ gravity. In sections (\textbf{2.1}) and (\textbf{2.2})
we reconstruct the ghost $f(R,T)$ model and discuss its stability for
homogeneous mater perturbation. Section \textbf{3} concludes our results.

\section{Reconstructing $f(R,T)$ Gravity}

The $f(R,T)$ gravity is an appealing modification to the
Einstein-Hilbert action by setting an arbitrary function of scalar
curvature $R$ and trace of the energy-momentum tensor $T$. The
action for this theory is defined as (Harko et al. 2011)
\begin{equation}\label{1}
\mathcal{I}=\int{dx^4\sqrt{-g}\left[\frac{1}{16\pi{G}}f(R,T)+\mathcal{L}_{(M)}\right]},
\end{equation}
where $\mathcal{L}_{(M)}$ denotes the matter Lagrangian.

The energy-momentum tensor of matter component is determined as
\begin{equation}\label{2}
T^{(M)}_{\alpha\beta}=-\frac{2}{\sqrt{-g}}\frac{\delta(\sqrt{-g}
{\mathcal{\mathcal{L}}_{(M)}})}{\delta{g^{\alpha\beta}}}.
\end{equation}
The corresponding field equations are found through the variation of
(\ref{1}) with respect to the metric tensor
\begin{eqnarray}\label{3}
&&\kappa^{2}T_{\alpha\beta}-f_{T}(R,T)T_{\alpha\beta}-f_{T}(R,T)
\Theta_{\alpha\beta}-R_{\alpha\beta}f_{R}(R,T)
+\frac{1}{2}g_{\alpha\beta}f(R,T)\nonumber\\&-&(g_{\alpha\beta}
{\Box}-{\nabla}_{\alpha}{\nabla}_{\beta})f_{R}(R,T)=0,
\end{eqnarray}
where
$f_{R}={\partial}f/{\partial}R,~f_{T}={\partial}f/{\partial}T,~
{\Box}={\nabla}_{\alpha}{\nabla}^{\beta};~{\nabla}_{\alpha}$ is the
covariant derivative linked with the Levi-Civita connection symbol
and $\Theta_{\alpha\beta}$ is defined by
\begin{equation}\label{4}
\Theta_{\alpha\beta}=\frac{g^{\mu\nu}{\delta}T_{\mu\nu}^{(M)}}
{{\delta}g^{\alpha\beta}}=-2T_{\alpha\beta}^{(M)}+g_{\alpha\beta}\mathcal{L}_M
-2g^{\mu\nu}\frac{\partial^2\mathcal{L}_M}{{\partial}
g^{\alpha\beta}{\partial}g^{\mu\nu}}.
\end{equation}
The modified field equation for the choice of perfect fluid are
given by
\begin{eqnarray}\label{5}
&&\kappa^{2}T_{\alpha\beta}+f_{T}(R,T)T_{\alpha\beta}+pf_{T}(R,T)
g_{\alpha\beta}-R_{\alpha\beta}f_{R}(R,T)+\frac{1}{2}g_{\alpha\beta}
f(R,T)\nonumber\\&-&(g_{\alpha\beta}{\Box}-{\nabla}_{\alpha}
{\nabla}_{\beta})f_{R}(R,T)=0.
\end{eqnarray}
In $f(R,T)$ gravity, the divergence of the energy-momentum tensor is
not covariantly conserved and is given by (Harko et al. 2011)
\begin{equation}\label{6}
\nabla^{\alpha}T_{\alpha\beta}=\frac{f_T}{\kappa^2-f_T}
\left[(T_{\alpha\beta}+\Theta_{\alpha\beta})
\nabla^{\alpha}\ln{f}_T+\nabla^{\alpha}\Theta_{\alpha\beta}-
\frac{1}{2}g_{\alpha\beta}\nabla^{\alpha}T\right].
\end{equation}
In this study, we consider the flat FRW geometry described by the
metric
\begin{equation*}
ds^{2}=dt^2-a^2(t)d\textbf{x}^2,
\end{equation*}
where $a(t)$ is the scale factor and $d\textbf{x}^2$ comprises the
spatial part of the metric. For the FRW metric with perfect fluid as
matter content, the divergence of the energy-momentum tensor takes
the form (Harko et al. 2011)
\begin{equation}\label{7}
\dot{\rho}+3H(\rho+p)=\frac{-1}{\kappa^2+f_T}\left[(\rho+p)\dot{T}f_{TT}+\dot{p}
f_T+\frac{1}{2}\dot{T}f_{T}\right].
\end{equation}
The above equation shows that energy momentum tensor is not
covariantly conserved due to the matter geometry coupling in this
theory. To obtain the standard continuity equation, we need to have
an additional constraint by taking the right side of the above
equation equal to zero. In this situation the additional constraint
is
\begin{equation}\label{8}
(1+\omega)Tf_{TT}+\frac{1}{2}(1-\omega)f_{T}=0.
\end{equation}
In next section, we consider particular $f(R,T)$ models and develop their
correspondence with the QCD GDE proposals.

\section{Ghost $f(R,T)$ Models}

Herein, we consider the $f(R,T)$ functions of the form
\begin{itemize}
  \item $f(R,T)=R+2f(T)$,
  \item $f(R,T)=f_1(R)+f_2(T)$.
\end{itemize}

\subsection{$f(R,T)=R+2f(T)$}

In the first place, we propose a particular case with $f(R,T)=R+2f(T)$. This
model corresponds to gravitational Lagrangian with time dependent
cosmological constant being function of trace of the energy-momentum tensor
(Poplawski 2006). Here, $f(T)$ is a correction to the usual Einstein Hilbert
action. Such model appears to be interesting and has been widely studied in
literature (Houndjo 2012; Sharif and Zubair 2013c, 2014a). The corresponding
field equations are
\begin{eqnarray}\label{9}
&&T_{\alpha\beta}+2f_{T}(T)T_{\alpha\beta}+2pf_T(T)-R_{\alpha\beta}+\frac{1}{2}g_{\alpha\beta}
(R+2f(T))=0.
\end{eqnarray}
The $00$ and $11$ component of filed equation (\ref{9}) can be
represented as
\begin{eqnarray}\label{10}
3H^2=\rho_M+\rho_{\vartheta}, \quad
-(2\dot{H}+3H^2)=p_{M}+p_{\vartheta},
\end{eqnarray}
where dot represents differentiation with respect to time,
$H=\dot{a}/{a}$ is the Hubble parameter and energy density
$\rho_{\vartheta}$ and pressure $p_{\vartheta}$ of \emph{dark energy
components} are obtained as
\begin{eqnarray}\label{11}
\rho_{\vartheta}&=&[2(\rho_M+p_M)f_T(T)+f(T)], \quad p_{dc}=-f(T).
\end{eqnarray}
The corresponding EoS parameter is
\begin{equation}\label{12}
\omega_{\vartheta}=\frac{-f(T)}{2(\rho_M+p_M)f_T(T)+f(T)}.
\end{equation}
\begin{itemize}
  \item Garcia-Salcedo GDE Model
\end{itemize}
Our aim is to reconstruct the $f(R,T)$ gravity according to QCD GDE model.
The QCD GDE energy density related to the dynamics of trapping horizon is
given by (Garcia-Salcedo et al. 2013)
\begin{eqnarray}\label{13}
&&\rho_{GDE}=\frac{\alpha(1-\epsilon)}{\tilde{r}_T}=\alpha(1-\epsilon)
\sqrt{H^2+\frac{\kappa}{a^2}}, \quad
\epsilon=\frac{\dot{\tilde{r}}_T}{2H\tilde{r}_T}.
\end{eqnarray}
For the flat case with spatial curvature $\kappa=0$, above relation
becomes
\begin{eqnarray}\label{14}
&&\rho_{GDE}=\alpha\left(1+\frac{\dot{H}}{2H^2}\right)H.
\end{eqnarray}
Using the energy conservation equation
$\dot{\rho_\vartheta}+3H\rho_\vartheta(1+\omega_\vartheta)=0$, the
EoS parameter is set of the form
\begin{eqnarray}\label{15}
1+\omega_{GDE}=-\frac{\dot{\rho}_{GDE}}{3H\rho_{GDE}}=\frac{1}{3}\left(\frac{\dot{\epsilon}}
{H(1-\epsilon)}+2\epsilon\right).
\end{eqnarray}
Equating the EoS parameters of dark energy components
$\omega_\vartheta$ and $\omega_{GDE}$ and hence using the constraint
(\ref{8}), it leads to
\begin{eqnarray}\label{16}
4T^2f_{TT}+\left[\left(1-\frac{1}{3}\left(\frac{\dot{\epsilon}}
{H(1-\epsilon)}+2\epsilon\right)\right)^{-1}-1\right]f=0.
\end{eqnarray}
It is evident that all the parameters in above equation are not
defined in terms of $T$ which makes it difficult to find the
analytic solution for $f(T)$. We are interested to determine $f(T)$
function coming from the QCD GDE model. In fact one can reconstruct
the actions in modified theories for known cosmic history in terms
of Hubble parameter. Here, we consider the Hubble parameter
presented in (Nojiri and Odintsov 2008)
\begin{eqnarray}\label{17}
H(t)=m(t_p-t)^{-\eta},
\end{eqnarray}
where $m$ and $\eta$ are positive constants and $t<t_p,~t_p$ is the
probable time when finite-time future singularity may appear. In
(Nojiri and Odintsov 2008), Nojiri et al. presented the
classification of finite future singularities. The $H(t)$ defined in
(\ref{17}) specifies these singularities as (Nojiri and Odintsov
2005, 2008; Bamba et al. 2010, 2012): type \textbf{I} (``Big rip
singularity") correspond to $\eta\geq1$, type \textbf{II} to the
$-1<\eta<0$, type \textbf{III} to the $0<\eta<1$ and type
\textbf{IV} singularity can appear for $\eta<-1$, but $\eta$ is not
an integer. Nojiri and Oditsov (2008) discussed the future evolution
of quintessence/phantom-dominated epoch in modified $f(R)$ gravity
which unifies the early-time inflation with late-time acceleration
inconsistent with observational tests. They discussed the models
where these singularities may occur. The occurrence of finite time
future singularities has also been studied in $f(T)$, modified
Gauss-Bonnet and $f(R,G)$ gravities (Bamba et al. 2010, 2012).
\begin{figure}
\centering \epsfig{file=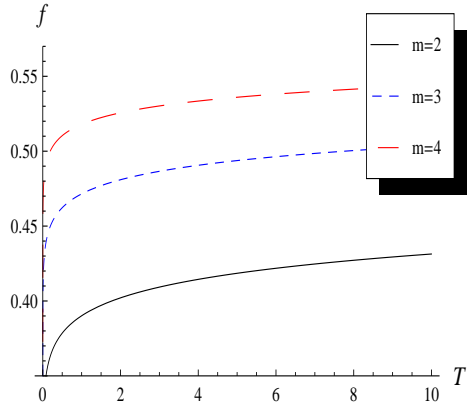, width=.495\linewidth,
height=2.2in} \caption{Evolution of $f(T)$ versus $T$ with
$H_0=67.3$ and $\Omega_{M0}=0.315$.}
\end{figure}
We are interested to discuss the specific case with $\eta=1$ and
$a(t)=a_0(t_p-t)^{-m},~a_0>0$. This model represents the phantom
phase of cosmos which may result in type \textbf{I} singularity. For
this choice of scale factor, the solution of Eq.(\ref{16}) is given
by
\begin{eqnarray}\label{18}
f(T)=c_1T^{\alpha_1}+c_2T^{\alpha_2},
\end{eqnarray}
where $\alpha_1=\frac{1}{2}(1-\sqrt{1-b})$,
$\alpha_2=\frac{1}{2}(1+\sqrt{1-b})$ with $b=1/(3m)$ and $c_1$, $c_2$ are
constants to be determined. Now evaluating the Friedmann equation (\ref{10})
at $t=t_0$ implies
\begin{equation}\label{19}
[1+2f_T(T_0)]\Omega_{M0}+\frac{f(T_0)}{3H_0^2}=1.
\end{equation}
After some manipulations it follows, it follows that
\begin{equation}\label{20}
f(T_0)=\frac{3H_0^2{\Omega}_{\vartheta0}}{b+1}.
\end{equation}
\begin{figure}
\centering \epsfig{file=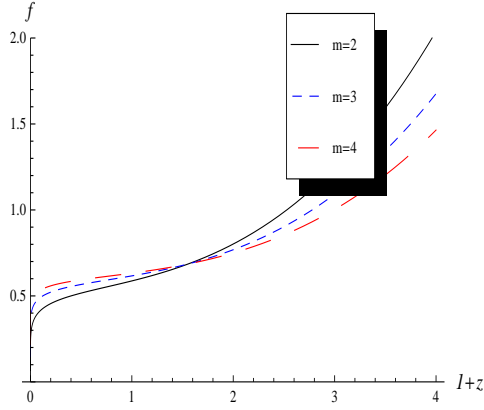, width=.495\linewidth,
height=2.2in} \caption{Evolution of $f(T)$ versus $z$ with
$H_0=67.3$ and $\Omega_{M0}=0.315$.}
\end{figure}
\begin{figure}
\centering \epsfig{file=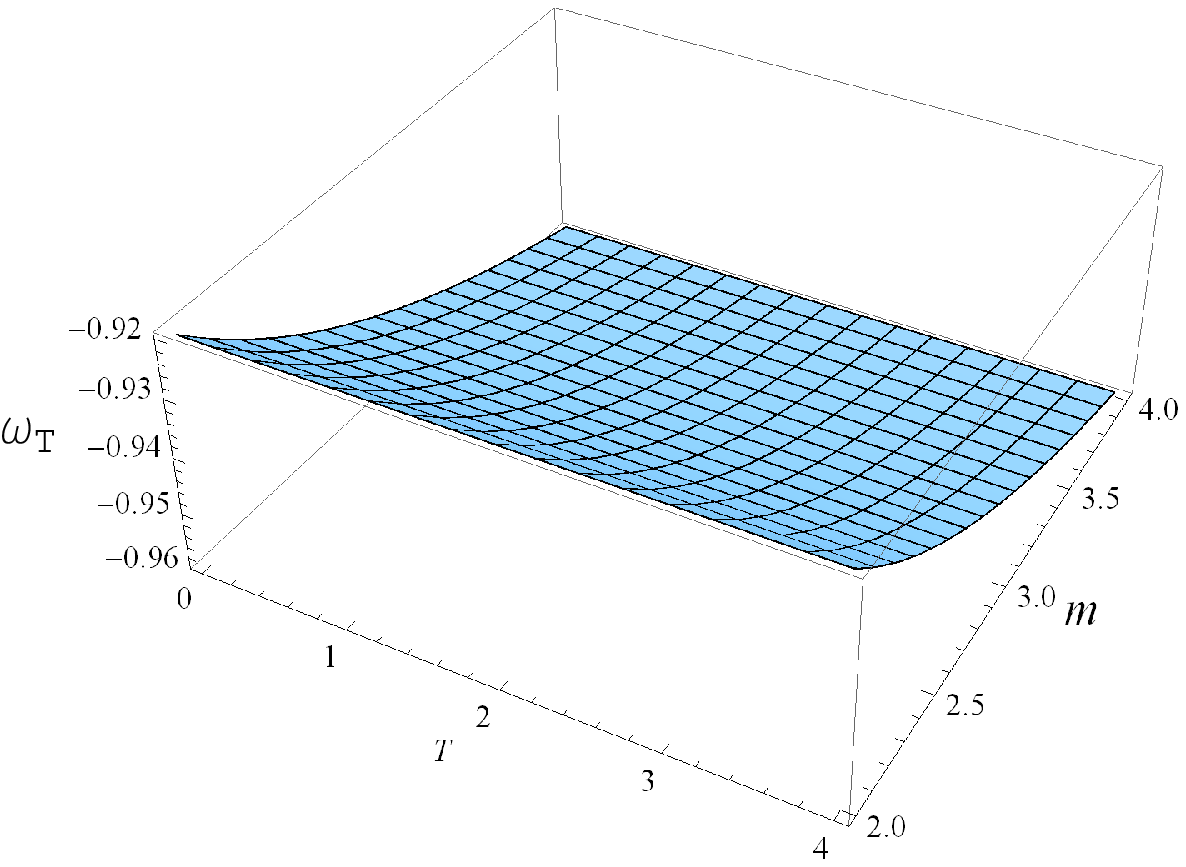, width=.495\linewidth,
height=2.2in} \caption{Evolution of EoS parameter for ghost $f(R,T)$
model with $H_0=67.3$ and $\Omega_{M0}=0.315$.}
\end{figure}
Hence using Eq.(\ref{20}) and the constraint (\ref{8}), we find the
constants $c_1$ and $c_2$ as
\begin{eqnarray}\nonumber
c_1=\left(\frac{3H_0^2{\Omega}_{\vartheta0}}{b+1}\right)\left(\frac{\alpha_2(1-2\alpha_2)}
{\alpha_1(2\alpha_1-1)-\alpha_2(2\alpha_2-1)}\right)T^{-\alpha_1}_0,\\\label{21}
c_2=\left(\frac{3H_0^2{\Omega}_{\vartheta0}}{b+1}\right)\left(\frac{\alpha_1(2\alpha_1-1)}
{\alpha_1(2\alpha_1-1)-\alpha_2(2\alpha_2-1)}\right)T^{-\alpha_2}_0.
\end{eqnarray}
We show the plot of $f(T)$ against $T$ for different values of $m$
in Figure 1. It can be seen that $f(T)$ increases depending on the
values of $T$, which is in accordance with the representation of the
function in Eq.(\ref{18}). We set the present day values of Hubble
parameter and fractional energy densities from the recent Planck
observations as $H_0=67.3$ and $\Omega_{M0}=0.315$ (Ade et al.
2013). The evolution of $f$ is also presented in terms of redshift
$z$ as shown in Figure 2. The behavior of equation of state
parameter $\omega_T$ in $f(R,T)$ gravity is shown in Figure 3.
$\omega_T$ helps to identify three significant eras of cosmic
expansion such as quintessence with $\omega_T>-1$, in this model
universe escapes from entering the de Sitter and big rip phases,
phantom $\omega_T<-1$, violating the null energy condition may
result in big rip phase and $\omega_T=-1$ results in de Sitter
phase. In our case it results in $\omega_T>-1$ representing the
quintessence era of the universe as shown in Figure 3.

\subsubsection{Stability of Ghost $R+f(T)$ model}

Here, we propose to explore the stability of ghost $f(R,T)$ model against
linear homogeneous perturbations. We assume a general solution $H(t)=H_h(t)$
for the dynamical equations in FRW background of $f(R,T)$ gravity. In
(Alvarenga et al. 2013b; Sharif and Zubair 2013d, 2014a), we have presented
the perturbed equations in the cosmological background of FRW universe for
both general as well as specific cases and discuss the stability of
$\Lambda$CDM, dS and power law solutions. We have also found the certain
constraints which have to be satisfied to ensure that power law solutions may
be stable and match the bounds prescribed by the energy conditions (Alvarenga
et al. 2013b; Sharif and Zubair 2013d, 2014a). The matter energy density in
terms of $H_h$ satisfies the relation
\begin{equation}\label{22}
\rho_h(t)=\rho_0 e^{-3\int{H_h(t)dt}},
\end{equation}
where $\rho_0$ is an integration constant. We propose to analyze the model
around the arbitrary solution $H_h(t)$, the Hubble parameter and energy
density can be perturbed as
\begin{equation}\label{23}
H(t)=H_h(t)(1+\delta(t)), \quad \quad \rho(t)=\rho_h
(1+\delta_m(t)).
\end{equation}
The function $f(T)$ can be expanded in powers of $T_h(=\rho_h$) as
\begin{equation}\label{24}
f(T)=f^h+f^h_T(T-T_h)+\mathcal{O}^2,
\end{equation}
where $\mathcal{O}^2$ term includes all the terms proportional to the squares
or higher powers of $T$ and the symbol $h$ means the functions and their
derivatives are evaluated corresponding to the solution $H(t)=H_h(t)$.

Using Eqs.(\ref{23}) and (\ref{24}) in FRW equation, we obtain
\begin{equation}\label{25}
(T_h+3T_hf^h_T+2T^2_hf_{TT}^h)\delta_m(t)=6H_h^2\delta(t).
\end{equation}
For the conserved energy momentum tensor the second perturbation equation is
\begin{equation}\label{26}
\dot{\delta}_m(t)+3H_h(t)\delta(t)=0.
\end{equation}
Combining Eqs.(\ref{25}) and (\ref{26}), we get the first order equation for
$\delta_m$ as
\begin{equation}\label{27}
\dot{\delta}_m(t)+\frac{1}{2H_h}(T_h+3T_hf_T^h+2T_h^2f_{TT}^h)\delta_m(t)=0.
\end{equation}
The evolution of $\delta_m$ and $\delta$ is determined by the relations
\begin{eqnarray}\nonumber
\delta_m(t)&=&\gamma\exp\left\{\frac{-1}{2}\int{\gamma_T}dt\right\},\quad
\delta(t)=\frac{\gamma{\gamma}_T}{6H_h}\exp\left\{\frac{-1}{2}\int{\gamma_T}dt\right\},\\\label{28}
\gamma_T&=&\frac{T_h}{H_h}(1+3f_T^h+2T_hf_{TT}^h).
\end{eqnarray}
Now, we examine the stability of ghost $f(R,T)$ model (\ref{18}). The
corresponding relations of $\gamma_T$ and $\frac{-1}{2}\int{\gamma_T}dt$ are
given as
\begin{eqnarray}\nonumber
\gamma_T&=&\frac{1}{m}\{T_0(t_p-t)^{3m+1}+\alpha(2\alpha+1)c_1T_0^\alpha(t_p-t)^{3\alpha{m}+1}
+\beta(2\beta\nonumber\\&+&1)c_1T_0^\beta(t_p-t)^{3\beta{m}+1}\},\\\nonumber
\frac{-1}{2}\int{\gamma_T}dt&=&\frac{1}{2m}\{\frac{T_0}{3m+2}(t_p-t)^{3m+2}
+\frac{\alpha(2\alpha+1)c_1}{3\alpha{m}+2}T_0^\alpha(t_p-t)^{3\alpha{m}+2}
\nonumber\\&+&\frac{\beta(2\beta+1)c_1}{3{\beta}m+2}T_0^\beta(t_p-t)^{3\beta{m}+2}\}.
\end{eqnarray}
It can be seen that the above expressions do not decay in future evolution
which results in increase of instability of our model against the homogeneous
perturbations. Thus the $f(R,T)$ ghost model is not stable in this scenario
which is in accordance with the results in (Cai et al. 2011). We also test
the evolution of squared speed of sound for stability analysis of ghost
$f(R,T)$ model. We plot $\nu_s^2$ against $T$ as shown in Figure 4. It can be
seen that squared speed of sound is negative showing the instability in the
model.
\begin{figure}
\centering \epsfig{file=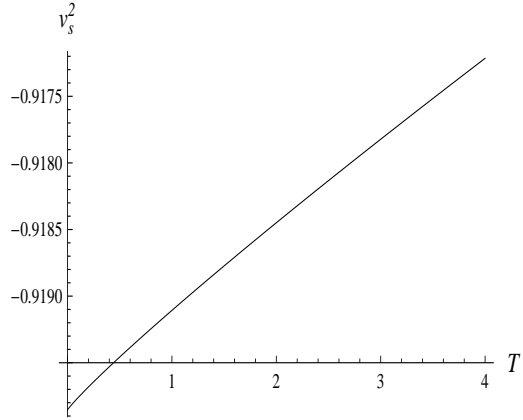, width=.495\linewidth,
height=2.2in} \caption{Evolution of $\nu_s^2$ versus $T$ with
$H_0=67.3$, $\Omega_{M0}=0.315$ and $m=2$.}
\end{figure}
\begin{itemize}
  \item Modified GDE Model
\end{itemize}
It can be seen that vacuum energy from the Veneziano ghost field in QCD turns
out to be $H+\mathcal{O}(H^2)$ (Zhitnitsky 2012). However, in previous works
of QCD GDE model people have only considered the case of
$\rho_{GDE}\propto{H}$. Recently, Cai et al. (2012) introduced a modified
form of QCD ghost model by involving the term $H^2$ so that energy density of
QCD ghost is given by
\begin{equation}\label{31}
\rho_{GDE}=\alpha{H}+\beta{H}^2,
\end{equation}
where $\alpha$ is same constant as that defined in the ordinary GDE and
$\beta$ is another constant with dimension $[energy]^2$. They fitted this
model  with observational data including SNeIa, BAO, BBN, the Hubble
parameter data and found that the modified GDE model, without having the two
fundamental cosmological puzzles, like the $\Lambda$CDM fit the astronomical
data very well. Sadeghi et al. (2013) discussed the flat universe model with
varying $G$ and $\Lambda$ in the presence of modified GDE model (\ref{31}).
In (Karami et al. 2013b), modified GDE scalar field models has been discussed
in FRW universe with both interaction and viscosity. The EoS parameter for
model (\ref{31}) is given by
\begin{eqnarray}\label{32}
1+\omega_{GDE}=-\frac{m(\alpha+2\beta{H})}{3(\alpha+\beta{H})}.
\end{eqnarray}
Comparing Eqs.(\ref{12}) and (\ref{32}), we find the $f(T)$ function of the
form
\begin{eqnarray}\nonumber
f(T)&=&C_1\exp\{-((2m-3)\beta+(m-3)\gamma){\ln}(-\alpha+\sqrt{\alpha^2+4T\gamma})+(m-3)
\\\nonumber&\times&(2\beta+\gamma) \ln(\alpha+\sqrt{\alpha^2+4T\gamma})+
\ln({\alpha(\beta+\gamma)+\beta\sqrt{\alpha^2+4T\gamma}})\\\label{33}&\times&3(\beta+\gamma)\}
/\{2m(2\beta+\gamma)\},
\end{eqnarray}
where $\gamma=3-\beta$ and $C_1$ is a constant to be determined. To find the
constant $C_1$, we set the initial condition by evaluating the Friedmann
equation (\ref{10}) at $t=t_0$ which implies
\begin{equation}\label{34}
f(T_0)=H_0^2\{2m\alpha\gamma+2m\beta(\alpha+\sqrt(\alpha^2+4\gamma{T_0}))\}/
\{2\alpha\gamma+\beta(\alpha+\sqrt(\alpha^2+4\gamma{T_0}))\}.
\end{equation}
Hence using (\ref{34}), constant $C_1$ is found of the form
\begin{eqnarray}\nonumber
C_1&=&H_0^2\{2m\alpha\gamma+2m\beta(\alpha+\sqrt(\alpha^2+4\gamma{T_0}))\}/
\{2\alpha\gamma+\beta(\alpha+\sqrt(\alpha^2+4\gamma{T_0}))\}
\\\nonumber&\times&
\exp\{-((2m-3)\beta+(m-3)\gamma){\ln}(-\alpha+\sqrt{\alpha^2+4T_0\gamma})+(m-3)
\\\nonumber&\times&(2\beta+\gamma) \ln(\alpha+\sqrt{\alpha^2+4T_0\gamma})+
\ln({\alpha(\beta+\gamma)+\beta\sqrt{\alpha^2+4T_0\gamma}})\\\nonumber&\times&3(\beta+\gamma)\}
/\{2m(2\beta+\gamma)\}.
\end{eqnarray}
In Figure 5(a), we plot the function $f(T)$ versus $T$ for different values
of $m$. Figure 5(b) shows the evolution of EoS parameter $\omega_{f(T)}$, and
it can be seen that $\omega_{f(T)}>0$. In this study we set the model
parameters according to Cai et al. 2012 results as $\alpha=2$, $\beta=0.3$
and $\gamma=3.342$. Hence this model is not viable according to recent
observations as it represents the matter dominated epoch. However, for
particular choice of $m$ it favors the quintessence era as shown in Figure 6.
\begin{figure}
\centering \epsfig{file=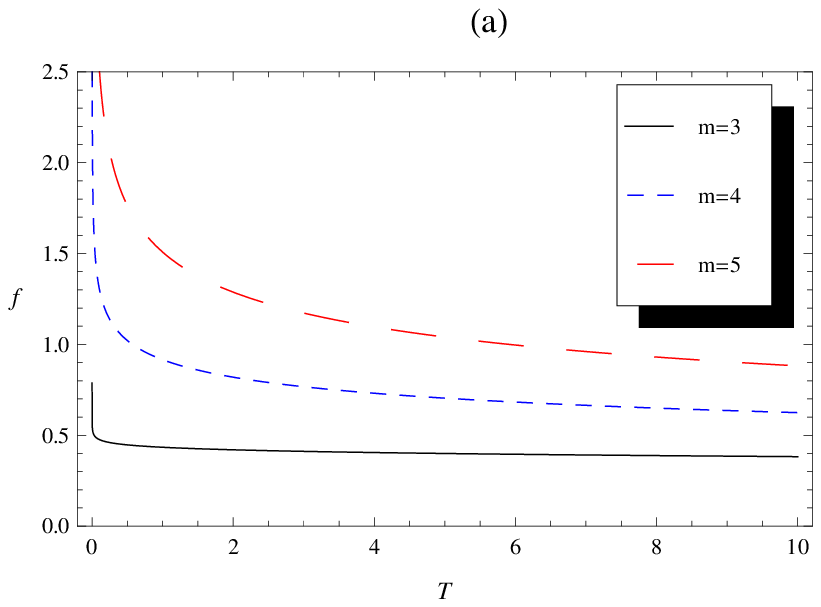, width=.49\linewidth,
height=2.1in}\epsfig{file=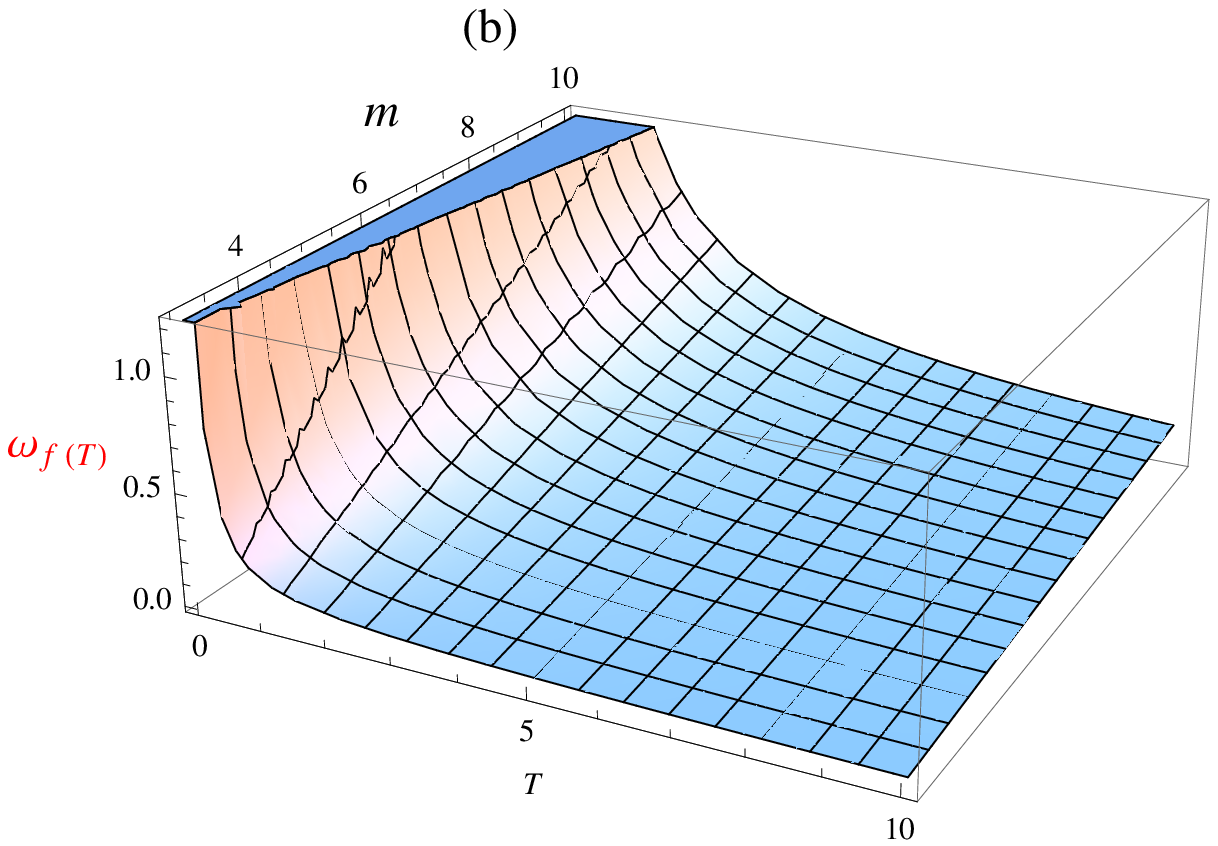, width=.49\linewidth,
height=2.1in} \caption{(a) Evolution of $f(T)$ versus $T$ with
$H_0=67.3$ and $\Omega_{M0}=0.315$. (b) Evolution of EoS parameter for ghost $f(R,T)$
model (\ref{33}).}
\end{figure}
\begin{figure}
\centering \epsfig{file=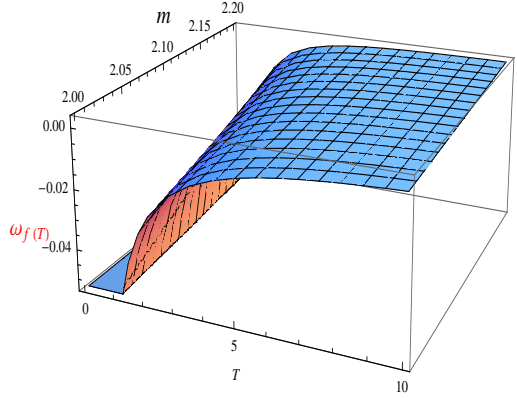, width=.49\linewidth,
height=2.1in} \caption{Evolution of EoS parameter for ghost $f(R,T)$
model (\ref{33}) with $H_0=67.3$ and $\Omega_{M0}=0.315$.}
\end{figure}

\subsection{$f(R,T)=f_1(R)+f_2(T)$}

In this section, we consider the Lagrangian as sum of two independent
functions of $R$ and $T$. Consequently, the corresponding field equations can
be arranged of the form
\begin{eqnarray}\label{40}
\dot{R}^2f_{1RRR}+(\ddot{R}-H\dot{R})f_{1RR}+[\kappa^2T+(1+\omega_{de})
\rho_{de}]f_{1R}+(\kappa^2+f_{2T})T=0,
\end{eqnarray}
which is the third order differential equation in $f_{1}$ and $f_2$ involving
contribution both from scalar curvature and matter density. To make the
Lagrangian $f(R,T)=f_1(R)+f_2(T)$ consistent with the standard continuity
equation, we need to set an additional constraint so that the right side of
the equation (\ref{7}). In such scenario, we have (Alvarenga et al. 2013a;
Sharif and Zubair 2014b)
\begin{equation*}
(1+\omega)Tf_{2TT}+\frac{1}{2}(1-\omega)f_{2T}=0,
\end{equation*}
which results in functional form of $f_2(T)$ as
\begin{equation}\label{41}
f_2(T)=\gamma_1T^\frac{1+3\omega}{2(1+\omega)}+\gamma_2,
\end{equation}
where $\gamma_i$'s are integration constants.
\begin{itemize}
  \item Garcia-Salcedo GDE Model
\end{itemize}
Initially, we consider the GDE model proposed by Garcia-Salcedo et al.
(2013). Using the GDE model (\ref{14}), matter energy density $\rho_m$ and
$\rho_{GDE}(1+\omega_{GDE})$ are represented in terms of Ricci scalar $R$ as
\begin{eqnarray}\nonumber
\rho_{GDE}(1+\omega_{GDE})&=&-\frac{\alpha\sqrt{2m+1}\sqrt{-R}}{(6m)^{3/2}},\\\label{42}
\rho_m&=&-\frac{mR}{2(2m+1)}\left(1-\frac{\alpha(2m+1)\sqrt{6m(2m+1)}}{6m^2\sqrt{-R}}\right).
\end{eqnarray}
Substituting the above results (\ref{41}) and (\ref{42}) in Eq.(\ref{40}), we
obtain a 3rd order nonlinear differential equation in terms of $f_1$ as
\begin{eqnarray}\nonumber
&&R^3f_{RRR}+\frac{3-m}{2}R^2f_{RR}+\left(\frac{3m^2\sqrt{-R}}{4}-\frac{\alpha(3m-1)(2m+1)^{3/2}}{4\sqrt{6m}}\right)f_R
\\\nonumber&+&\frac{1}{8}(6m^2R-\alpha(2m+1)^{3/2}\sqrt{-6mR})-\frac{3\gamma_1\sqrt{-(2m+1)R}}{4\sqrt{2}}
\\\label{43}&\times&\sqrt{1-\frac{\alpha(2m+1)\sqrt{6m(2m+1)}}{6m^2\sqrt{-R}}}=0.
\end{eqnarray}
The analytic solution of above equation is not on cards, therefore we solve
this equation numerically by setting the following initial conditions
(Capozziello et al. 2005; Houndjo 2012; Sharif and Zubair 2014b)
\begin{eqnarray}\nonumber
&&f_R\mid_{t=t_0}=1, \quad f_{RR}\mid_{t=t_0}=0,\\\nonumber
&&f(R_0)=R_0+\delta,\quad \delta=6H_0^2(1-\Omega_{m0}-\frac{\sqrt{3\Omega_{m0}}\gamma_1}{2H_0})
-\frac{\gamma_2}{3H_0^2}.
\end{eqnarray}
\begin{figure}
\centering \epsfig{file=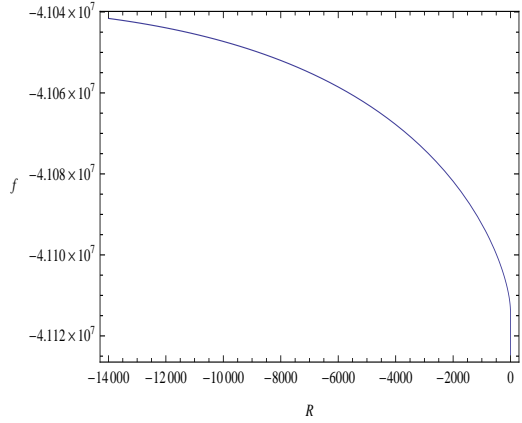, width=.495\linewidth,
height=2.2in} \caption{Evolution of $f$ versus $R$ with $H_0=67.3$,
$\Omega_{M0}=0.315$, $\alpha=-10^{-6}$, $\gamma_1=\gamma_2=1$ and
$m=2$.}
\end{figure}

In numerical solutions, we set the present day values of parameters as
$H_0=67.3$, $\Omega_{M0}=0.315$ and $\Omega_{DE}=0.685$ (Ade et al. 2013).
The variation of $f$ with $R$ is shown in Figure 7. In this plot, we set
$m=2$ while other values of $m$ do not make major changes on behavior of
curves as presented in Figure 7. The ($f-R$) plot drawn in Figure 7 provides
sufficient data set which can be used to find closed analytic mathematical
expression for $f$ in terms of $R$. These expressions can be determined by
using a polynomial function for $f$ as $f={\Sigma^{n}_{0}}a_{i}R^{i}$. The
approximate analytic function corresponding to Figure 7 may be expressed as
\begin{eqnarray}\nonumber
f(R)&=&-4.112\times10^7-127.768R-0.502R^2-0.001R^3-1.759\times10^{-6}R^4\label{44}\\&-&1.432\times10^{-9}R^5
-6.192\times10^{-13}R^6-1.102\times10^{-16}R^7.
\end{eqnarray}
\begin{figure}
\centering \epsfig{file=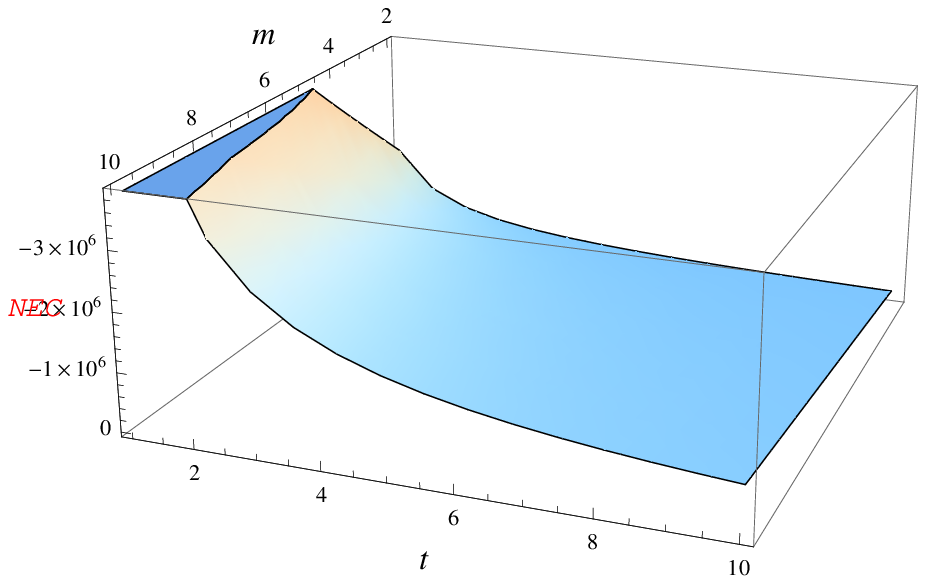, width=.495\linewidth,
height=1.8in}\epsfig{file=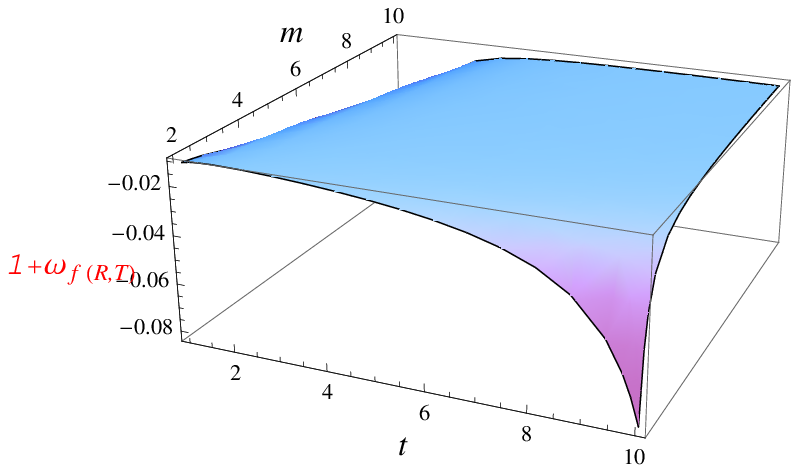, width=.495\linewidth,
height=1.8in} \caption{(a) Evolution of NEC and (b) EoS parameter
$\omega_{f(R,T)}$ for ghost $f(R,T)$ model (\ref{44}) with
$H_0=67.3$, $\Omega_{M0}=0.315$, $\alpha=-10^{-6}$,
$\gamma_1=\gamma_2=1$ and $m=2$.}
\end{figure}
In further study, we use Eq.(\ref{44}) to show the evolution of EoS parameter
and null energy condition for $f(R,T)=f_1(R)+f_2(T)$ ghost DE model. Figure
8(a) shows the variation of NEC versus power law parameter $m$ and time $t$.
It can be seen that NEC is violated \emph{i.e.}, $\rho_T+p_T<0$ which
necessitates $\omega_T<-1$ (the phantom regime of the cosmos). To see this
behavior of $f(R,T)$ model (\ref{44}), we plot the evolution of $\omega_T<-1$
versus $t$ and $m$ as shown in Figure 8(b).

The energy density $\rho_\vartheta$ and pressure $p_\vartheta$ for the
$f(R,T)$ model, are defined as
\begin{eqnarray}\nonumber
\rho_{\vartheta}&=&\frac{1}{f_{1R}}\left[(1-f_{1R})\rho+\alpha_1\sqrt{T}+\alpha_2
+\frac{1}{2}(f_1-Rf_{1R}) -3H\dot{R}f_{1RR}\right],\\\nonumber
p_{\vartheta}&=&\frac{1}{f_{1R}}
\left[\frac{1}{2}(Rf_{1R}-f_1)-\frac{\alpha_1}{2}\sqrt{T}+\alpha_2+(\ddot{R}+2H\dot{R})f_{1RR}+\dot{R}^2f_{1RRR}\right].
\end{eqnarray}
Using the above relations of $\rho_\vartheta$ and $p_\vartheta$, we define
the squared sound speed of GDE
$v_s^2=\dot{p_\vartheta}/{\dot{\rho_\vartheta}}$. In plot 9, we show the
evolution of $v_s^2$ for the $f(R,T)$ model (\ref{44}). It can be seen that
$v_s^2$ is less than zero.
\begin{figure}
\centering \epsfig{file=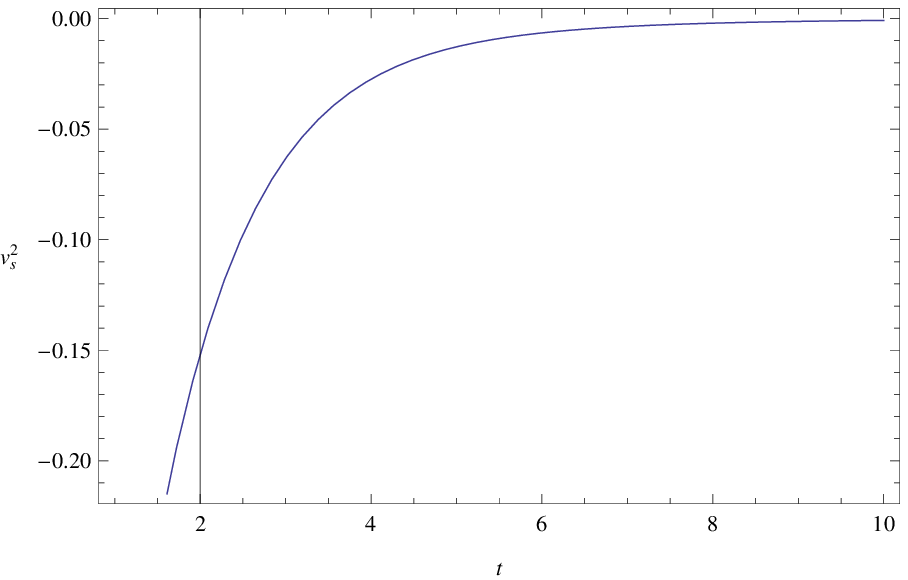, width=.495\linewidth,
height=2.2in} \caption{Evolution of $v_s^2$ versus $t$ with
$H_0=67.3$, $\Omega_{M0}=0.315$, $\alpha=-10^{-6}$,
$\gamma_1=\gamma_2=1$ and $m=2$.}
\end{figure}
\begin{itemize}
  \item Modified GDE Model
\end{itemize}
Here, we reconstruct the $f(R,T)=f_1(R)+f_2(T)$ function corresponding to
modified GDE model (Cai et al. 2012). Using the model (\ref{31}), we set
expressions for $\rho_m$ and $\rho_{GDE}(1+\omega_{GDE})$ as given below
\begin{eqnarray}\label{45}
\rho_m&=&\frac{-mR}{2(2m+1)}\left(1-\frac{1}{3}\left(\beta+\frac{\alpha\sqrt{6(2m+1)}}{\sqrt{-mR}}\right)\right),\\\label{46}
\rho_\vartheta(1+\omega_\vartheta)&=&\frac{-1}{18(2m+1)}\left(\alpha\sqrt{-6m(2m+1)R}-2\beta{R}\right).
\end{eqnarray}
Using Eqs.(\ref{45}) and (\ref{46}), we finally find the following results
\begin{eqnarray}\nonumber
&&R^3f_{RRR}+\frac{3-m}{2}R^2f_{RR}-\left(\frac{3m^2R}{2}\left(1-\frac{1}{3}\left(\beta+\frac{\alpha\sqrt{6(2m+1)}}{\sqrt{-mR}}\right)\right)
\right.\nonumber\\&+&\left.\frac{m}{12}\left(\alpha\sqrt{-6m(2m+1)R}-2\beta{R}\right)\right)f_R+\frac{3m^2R}{4}
\left(1-\frac{1}{3}\left(\frac{\alpha\sqrt{6(2m+1)}}{\sqrt{-mR}}\right.\right.\nonumber\\&+&\left.\left.\beta\right)\right)
-\frac{3m(2m+1)\alpha_1}{4}\sqrt{\frac{-mR}{2(2m+1)\left(1-\frac{1}{3}\left(\beta+\frac{\alpha\sqrt{6(2m+1)}}{\sqrt{-mR}}\right)\right)}}=0,
\end{eqnarray}
which is a 3rd order nonlinear differential equation. Again we solve this
equation numerically using the initial conditions as in previous case. We
show the variation of $f$ with respect to $R$ in Figure 10 with the choice of
parameters $H_0=67.3$, $\Omega_{M0}=0.315$, $\Omega_{DE}=0.685$ (Ade et al.
2013), $\alpha=2$, $\beta=0.3$ and $m=2$.
\begin{figure}
\centering \epsfig{file=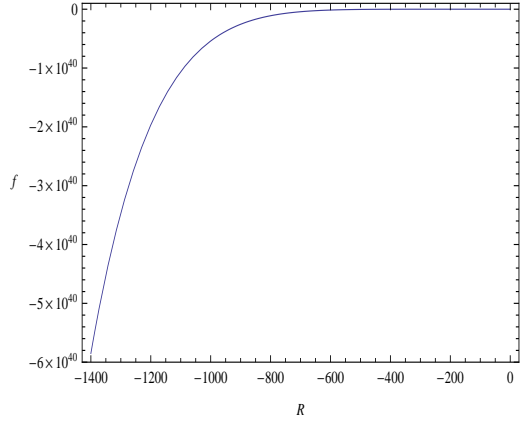, width=.495\linewidth,
height=2.2in} \caption{Variation of $f$ versus $R$ for modified GDE
model with $H_0=67.3$, $\Omega_{M0}=0.315$, $\alpha=2$, $\beta=0.3$,
$\gamma_1=\gamma_2=1$ and $m=2$.}
\end{figure}
One can find the analytic function corresponding to Figure 10 so that
\begin{eqnarray}\nonumber
&&f(R)=8.139\times10^{32}+4.079\times10^{33}R+4.912\times10^{31}R^2+2.207\times10^{29}R^3
\label{48}\\&+&4.836\times10^{26}R^4+5.716\times10^{23}R^5+8.577\times10^{20}R^6+5.999\times10^{18}R^7.
\end{eqnarray}
Using Eq.(\ref{48}), we discuss the variation of NEC and also EoS parameter
for the obtained $f(R,T)$ model. In Figure 11, we show evolution of NEC and
EoS parameter versus $m$ and $t$. It can be seen that NEC is violated for
$m\geq60$ as shown in Figure 11(a) and EoS parameter favors the phantom
regime for this choice of $m$. The evolution of squared speed of sound is
presented in Figure 12.
\begin{figure}
\centering \epsfig{file=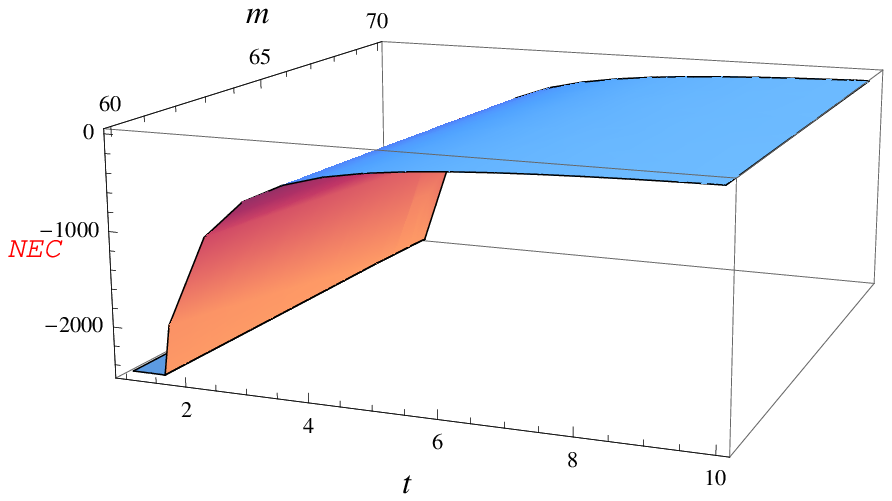, width=.495\linewidth,
height=1.8in}\epsfig{file=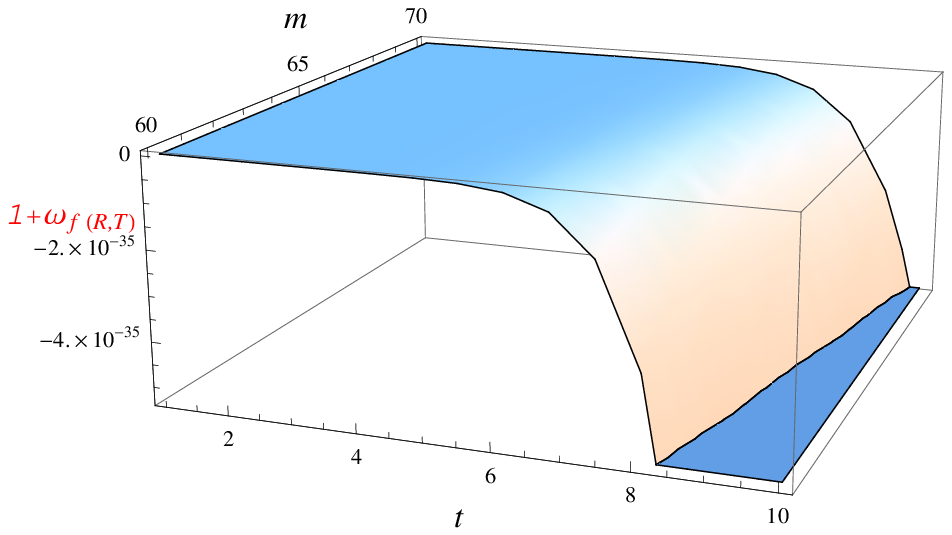, width=.495\linewidth,
height=1.8in} \caption{(a) Evolution of NEC and (b) EoS parameter $\omega_{f(R,T)}$ for modified ghost $f(R,T)$
model (\ref{48}) with $H_0=67.3$, $\Omega_{M0}=0.315$, $\alpha=2$, $\beta=0.3$ and $\gamma_1=\gamma_2=1$.}
\end{figure}
\begin{figure}
\centering \epsfig{file=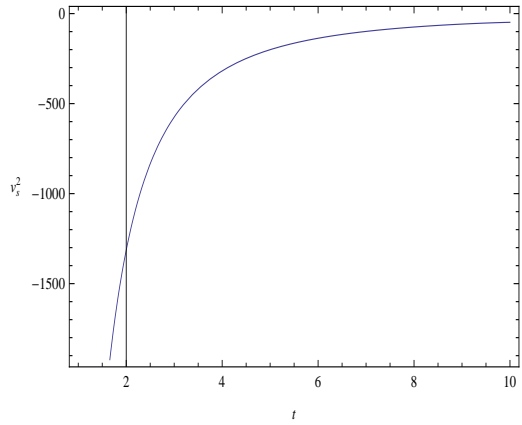, width=.495\linewidth,
height=2.2in} \caption{Evolution of $v_s^2$ for modified ghost $f(R,T)$
model (\ref{48}) versus $t$ with $H_0=67.3$, $\Omega_{M0}=0.315$, $\alpha=2$, $\beta=0.3$, $\gamma_1=\gamma_2=1$ and $m=2$.}
\end{figure}

\section{Conclusions}

Modified theories of gravity have appeared as convenient candidates
to address issues of accelerated cosmic expansion and predict the
destiny of the universe. Harko et al. (2011) generalized $f(R)$
gravity by introducing an arbitrary function of the Ricci scalar $R$
and the trace of the energy-momentum tensor $T$. The dependence of
$T$ may be introduced by exotic imperfect fluids or quantum effects
(conformal anomaly). $f(R,T)$ is a general modified gravity
formulated on the basis of curvature matter coupling and provides an
alternative way to explain the current cosmic acceleration with no
need of introducing either the existence of extra spatial dimension
or an exotic component of dark energy.

$f(R,T)$ gravity has gained significant attention to handle the issue of
accelerated cosmic expansion and various aspects have been explored. In this
respect the search of most appropriate form of Lagrangian is still under
consideration. $f(R,T)$ theory of gravity has been reconstructed under
various scenarios including de Sitter, power law solutions, phantom,
non-phantom eras, anisotropic universe model and class of HDE models (Houndjo
2012; Sharif and Zubair 2013c, 2014a). In this work, we reconstruct $f(R,T)$
gravity corresponding to Garcia-Salcedo and modified QCD ghost DE models. We
consider the $f(R.T)$ models of the form $f(R,T)=R+2f(T)$ and
$f(R,T)=f_1(R)+f_2(T)$. A particular model of scale factor is considered
representing the phantom phase of the cosmos which may result in type
\textbf{I} singularity (Nojiri and Odintsov 2008). The major concern of
theories involving non-minimal matter geometry coupling is the divergence of
energy momentum tensor is not covariantly conserved. We have obtained the
explicit form of functions $f(T)$ and $f(R)$ using the constraint of standard
continuity equation. In the following we summarize our findings
\begin{itemize}
  \item $f(R,T)=R+2f(T)$
\end{itemize}
In the first place we consider a $f(R,T)$ model representing a
correction to Einstein gravity in the form of time dependent
cosmological constant. We have selected two generalized form of GDE
models, one suggested by Garcia-Salcedo et al. (2013) and other
model is suggested by Cai et al. (2012) named as MGDE. The $f(T)$
model corresponding to Garcia-Salcedo GDE is introduced in
Eq.(\ref{18}). The evolution of $f(T)$ and EoS parameter is shown in
Figures 1-3. Figure 3 shows that this model represents the
quintessence era of the universe which is consistent with the WMAP9
observations $-1.71<\omega<-0.34$ (Hinshaw et al. 2013). We examine
the stability of ghost $f(R,T)$ model against linear homogeneous
perturbations. We find that ghost $f(R,T)$ model is not stable in
this scenario. We also test the squared speed of sound $\nu_s^2$ and
plot its evolution in Figure 4. It shows the instability of the
ghost $f(R,T)$ model. In Section 3.1, the reconstruction of $f(T)$
function is also developed corresponding to MGDE model. The
evolution of $f(T)$ versus $T$ and $f(R,T)$ EoS parameter for model
(\ref{33}) is shown in Figure 5. We find that this model is not
viable according to recent observations.
\begin{itemize}
  \item $f(R,T)=f_1(R)+f_2(T)$
\end{itemize}
In this scheme $f_2(T)$ is found using the constraint for
conservation equation in $f(R,T)$ gravity. Substituting this
$f_2(T)$ function and parameters for the particular GDE models in
dynamical equations of $f(R,T)$ gravity results in 3rd order
nonlinear equations in terms of $f_1$. We solve such type of
equations using numerical technique and find the approximate
analytic function corresponding to plots of resulting functions. In
case of Garcia-Salcedo $f(R,T)$ model, we show the evolution of NEC
and EoS parameter in Figure 8. We find that $\omega$ favors the
phantom regime of the universe and lies in the range
$-1.08<\omega<-1.02$ which is consistent with ranges set by Planck
and WMAP9 data (Ade et al. 2013; Hinshaw et al. 2013) in the
following form
\begin{itemize}
  \item $\omega=-1.13^{+0.13}_{-0.14}$, (Planck+WP+SNLS)
  \item $\omega=-1.084\pm{0.0063}$, (WMAP+eCMB+BAO+$H_0$+SNe)
\end{itemize}
In case of MGDE DE candidate, we set the model parameters following
(Cai et al. 2012) and show the variation of NEC and EoS parameter in
Figure 11. It is found that $\omega$ lies in the range
$-6\times10^{-35}<1+\omega<-1$ representing the phantom era of the
universe consistent with the recent observations (Ade et al. 2013;
Hinshaw et al. 2013).

\section{Conflict of Interest}

The authors declare that they have no conflict of interest.

\vspace{.25cm}

{\bf Acknowledgment}

\vspace{.25cm}

{The authors acknowledges the anonymous referee for enlightening comments
that helped to improve the quality of the manuscript.}

\end{document}